\documentclass[12pt,epsf]{article}
\usepackage{amsmath,amssymb,graphicx,epsfig,latexsym}
\numberwithin{equation}{section}
\usepackage{feynmf}

\newcommand{\be}{\begin{equation}}
\newcommand{\ee}{\end{equation}}

\newcommand{\beqa}{\begin{eqnarray}}
\newcommand{\eeqa}{\end{eqnarray}}
\newcommand{\nn}{\nonumber}

\def\boxit#1{\vbox{\hrule\hbox{\vrule\kern8pt
\vbox{\hbox{\kern8pt}\hbox{\vbox{#1}}\hbox{\kern8pt}}
\kern8pt\vrule}\hrule}}
\def\mathboxit#1{\vbox{\hrule\hbox{\vrule\kern8pt\vbox{\kern8pt
\hbox{$\displaystyle #1$}\kern8pt}\kern8pt\vrule}\hrule}}

\def\IB{\relax\hbox{$\inbar\kern-.3em{\rm B}$}}
\def\IC{\relax\hbox{$\inbar\kern-.3em{\rm C}$}}
\def\ID{\relax\hbox{$\inbar\kern-.3em{\rm D}$}}
\def\IE{\relax\hbox{$\inbar\kern-.3em{\rm E}$}}
\def\IF{\relax\hbox{$\inbar\kern-.3em{\rm F}$}}
\def\IG{\relax\hbox{$\inbar\kern-.3em{\rm G}$}}
\def\IGa{\relax\hbox{${\rm I}\kern-.18em\Gamma$}}
\def\IH{\relax{\rm I\kern-.18em H}}
\def\IK{\relax{\rm I\kern-.18em K}}
\def\IL{\relax{\rm I\kern-.18em L}}
\def\IP{\relax{\rm I\kern-.18em P}}
\def\IR{\relax{\rm I\kern-.18em R}}
\def\IZ{\relax\ifmmode\mathchoice
{\hbox{\cmss Z\kern-.4em Z}}{\hbox{\cmss Z\kern-.4em Z}}
{\lower.9pt\hbox{\cmsss Z\kern-.4em Z}} {\lower1.2pt\hbox{\cmsss
Z\kern-.4em Z}}\else{\cmss Z\kern-.4em Z}\fi}

\def\II{\relax{\rm I\kern-.18em I}}

\def\CO {{\cal O}}

\begin{document}
\setlength{\unitlength}{1mm}
\setlength{\baselineskip}{7mm}
\begin{titlepage}

\begin{flushright}

{\tt NRCPS-HE-14-2015} \\

\end{flushright}

\vspace{1cm}
%\begin{titlepage}
%\title{
\begin{center}

{\Large \it  Proton Spin  and Tensorgluons
\\
\vspace{0,3cm}

} %title ends

\vspace{1cm}
%\author{
{\sl Spyros Konitopoulos
 and  George Savvidy

\bigskip
\centerline{${}^+$ \sl Demokritos National Research Center, Ag. Paraskevi,  Athens, Greece}
\bigskip
}%author ends
%}
%\date{}%in order NOT to write the date
%\maketitle
\end{center}
\vspace{25pt}

\centerline{{\bf Abstract}}

Recently it was suggested that inside the hadrons there are additional
partons - tensorgluons - which carry the same charges as the gluons, but have larger spin.
The nonzero density of tensorgluons can be created inside a nucleon by radiation
of tensorgluons by gluons.   Tensorgluons can carry a part of nucleon momentum
together with gluons.  Because tensorgluons have a larger spin,  they
can influence the spin structure of the nucleon.
We analyse a possible contribution of polarised tensorgluon density
to the spin of the nucleon.
This contribution appears in the next to leading order, compared to the gluons
and can provide a substantial screening effect due to the larger spin of tensorgluons.

\vspace{10pt}

\noindent

\begin{center}
%Keywords:~~ Gauge Fields; Tensor Gauge Fields;
 
\end{center}

\vspace{150 pt}

%\end{abstract}
%\thispagestyle{empty}
%\end{titlepage}

\end{titlepage}

\newpage

\pagestyle{plain}
%\pagenumbering{roman}

\section{\it Introduction}

In recent articles \cite{Savvidy:2014hha,Savvidy:2013zwa} it was assumed that inside a nucleon there are additional
partons - tensorgluons - which carry the same charges as the gluons, but have larger spins.
The nonzero density of tensorgluons can be created inside a nucleon by radiation
of tensorgluons by gluons.  The tensorgluons can carry a part of nucleon momentum
together with gluons.  Because tensorgluons have larger a spin s (s=2,3,...), they
can influence the spin structure of the nucleon.
Our aim is to analyse a possible contribution of polarised tensorgluon density
$\Delta T_s$ into the spin of the nucleon, if they are present in the nucleon.

The spin structure of the nucleon remains the essential problem of nonperturbative QCD
and hadronic physics. One of its most significant manifestations is the so-called
spin crisis,
or spin puzzle,  related to the surprisingly small fraction of proton polarisation carried by the
quarks \cite{Ashman:1989ig,Anthony:2000fn,Alekseev:2010hc,Adolph:2012ca, 
Airapetian:2006vy,deFlorian:2009vb,Anselmino:1994gn,Aidala:2012mv}.
The contribution from the quarks spin is found out to be small, approximately  25\% of
the total proton spin,  and it is expected that the rest should come from the
spin of gluons, quark sea polarisation and the
orbital angular momentum of quarks and gluons.
This problem attracted the attention to the peculiarities of the underlying QCD
description of the nucleon spin, and, in particular, to the role of the gluons
\cite{Kodaira:1978sh,Kodaira:1979ib,Kodaira:1979pa,Jaffe:1987sx,Efremov:1988zh,
Altarelli:1988nr,Dokshitzer:1991wu}.
One of the most interesting suggestions to come out of the spin crisis
was that gluon polarisation $\triangle G$ may contribute significantly to the nucleon
spin \cite{Jaffe:1987sx,Efremov:1988zh,Altarelli:1988nr,Carlitz:1988ab}.
The amount of gluon polarisation from the experiment is approximately $20\%$
\cite{Adolph:2012ca,Jimenez-Delgado:2013sma,Mallot:2006mc,Nocera:2012hx,Ball:2013lla}
and it is not by itself sufficient to resolve the problem.
It remains to identify the rest ($ 50\%$ )  of the nucleon spin.
The experimental indications
of the small value of the polarisation carried by the
quarks and gluons point to a possible contribution of the orbital angular momentum of
the constituent patrons  or {\it to a possible  contribution coming   from the polarisation of additional
kind of partons in the nucleon}.

Following the suggestion \cite{Savvidy:2014hha,Savvidy:2013zwa}
that there are additional patrons in the nucleon,
one can propose  that the nucleon's spin is accounted
by the quark spin ${1\over 2} \Delta \Sigma$,   gluon spin $1 \Delta G$, as well as  by the tensorgluon spin  $s \Delta T_s$ and the total orbital angular momentum
$L_z$. Physically  the quantities  $\Delta \Sigma,\Delta G ,
 \Delta T_s $ represent the differences
between the quark number densities, gluons
and tensorgluons with spin parallel to the nucleon
spin  and those with spin anti-parallel (\ref{poladistributions}).
Because the spin of the nucleon is ${1\over 2} \hbar$, one should have the following
spin sum rule of helicity weighted distributions:
\be\label{spinsumrule}
{1\over 2}  \Delta \Sigma + \Delta G + \sum_{s} s ~\Delta T_s  + L_z  ={1\over 2}.
\ee
In this equation the helicity of the gluon is equal to one, of the tensorgluons
is $s=2,3,..$ and
the summation is over all "active" tensorgluons in the nucleon.
The various components of the nucleon spin in the above
equation can be measured experimentally
\cite{Adolph:2012ca,deFlorian:2009vb,Anselmino:1994gn,Aidala:2012mv,
Mallot:2006mc,Ellis:2005cy,Nocera:2012hx,Ball:2013lla}.  
Our aim is to consider a possible contribution
of tensorgluons of spin s, the  $s \Delta T_s$ contribution.

 \begin{figure}
 \includegraphics[width=8cm]{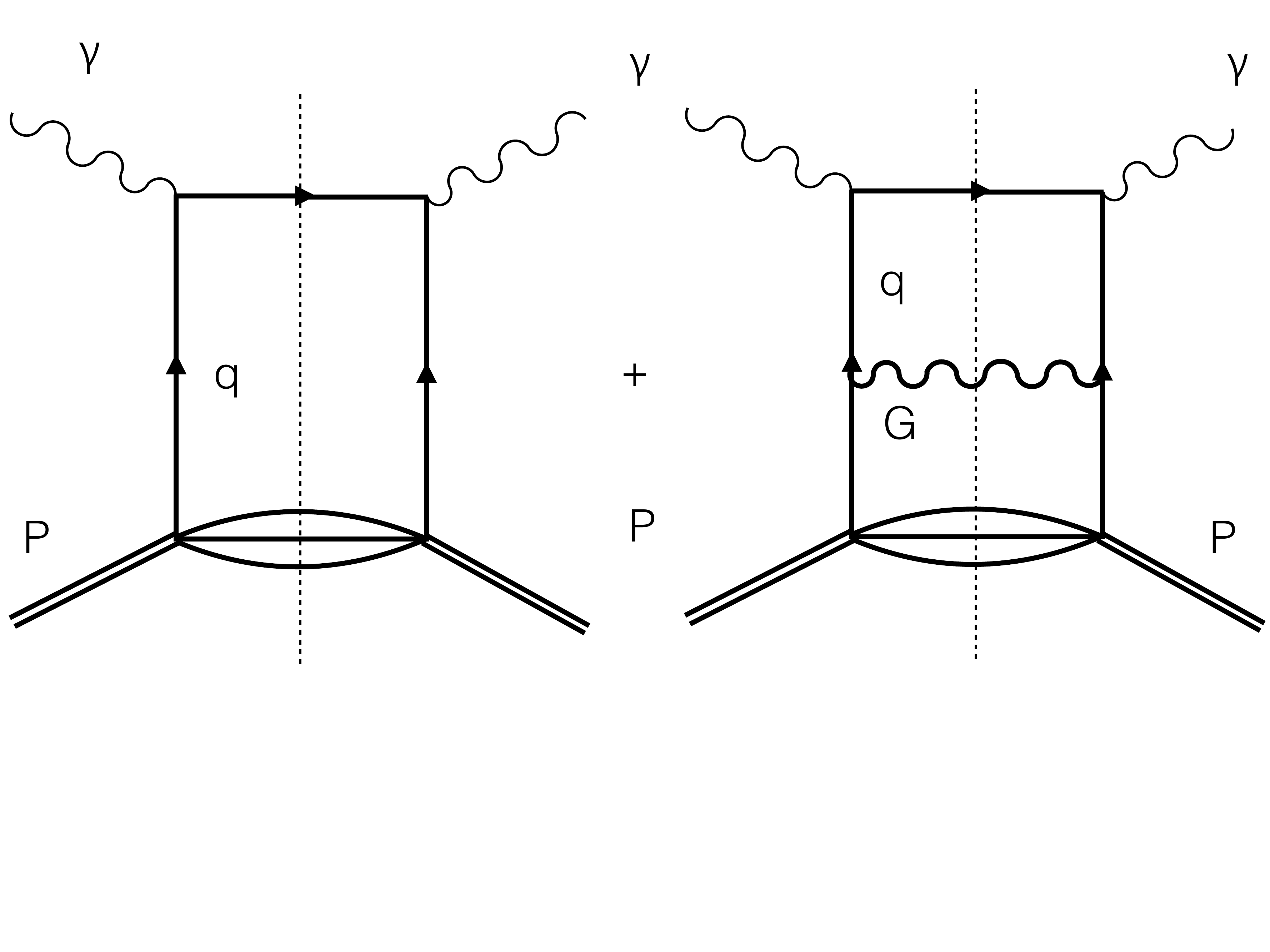}
 \centering
 \caption{Scattering of a photon on a proton quark q in the leading  and next to leading
 order through the gluon exchange.  }
 \label{fig1}
 \end{figure}

Considering the influence of polarised tensorgluons density
$\Delta T_s$ to the  singlet part of the first moment $g_1$ in  (\ref{moments1})
we found  an additional contribution from nucleon  tensorgluons:
\beqa\label{totalgluontensorcontribition}
 I_0&=&{1\over 9 }~(\Delta \Sigma - n_f {\alpha(Q^2) \over 2 \pi} \Delta G )~
 \left(1-{\alpha(Q^2)\over 2 \pi}
 {3 (12s^2-1)-8n_f\over 3 (12s^2-1)-2n_f} \right) +\nn\\
 &&~~~~~+   n_f    {\alpha (Q^2)\over  2 \pi}    \Delta T_s ~
{ \sum_{k=1}^{2s+1}{1\over k}\over 3 (12s^2-1)-2n_f} ~.
\eeqa

The present paper is organized as follows. In section 2 the basic formulae for polarised electroproduction are recalled and short review of the experimental data is presented.
In section 3 we are presenting the evolution equations that
describe the $Q^2$ dependence of polarised parton densities including
polarised tensorgluons.  In section 4 the solution of the evolution equations are considered
in leading and next to leading order.   In Appendix we present the
polarised splitting functions including tensorgluons and the corresponding
anomalous dimensions \cite{Savvidy:2014hha,Savvidy:2013zwa}.

\section{\it The Polarised Electroproduction}

The polarised electroproduction is described by two structure functions $G_1$ and
$G_2$ in addition to the unpolarised structure functions $F_1$ and
$F_2$. It is convenient to use the scaling functions  $g_1(x,Q^2)= M (p\cdot q) G_1$ and
$g_2(x,Q^2)= (p\cdot q)^2 G_2 /M $, where p and M are the momentum
and mass of the nucleon,    $q$ is the momentum  of
the virtual photon with  $Q^2 =-q^2$ and $x = Q^2/ 2 (p\cdot q)$.
These spin-dependent structure functions $g_i$ can be extracted from measurements
where longitudinally polarised leptons are scattered from a target that is polarised either
longitudinally or transversely relative to the electron beam \cite{Aidala:2012mv}.
For longitudinal beam and target
polarisation, the difference between the cross sections for spins aligned and antialigned is
dominated at high energy by the $g_1$ structure function. The $g_2$ structure function can be
determined with additional measurement of cross sections for a nucleon polarised in a direction
transverse to the beam polarisation. The moments
of the spin structure functions are defined as
\be\label{structuremoments}
\int^{1}_{0} dx ~x^{n-1} ~g_{i}(x,Q^2),~~~~i=1,2.
\ee
In analogy with the unpolarised $F_1$ structure function
\be
 F_1 = {1\over 2 }  \sum_{i}  e^{2}_{i} [  q^{i}(x,Q^2) +  \bar{q}^{i}(x,Q^2) ]
\ee
the structure function $g_1$ can be expressed at leading order of perturbation theory in
terms of differences between quark distributions with spins aligned $q^{i}_{+}$
and antialigned $q^{i}_{-}$ relative to that of the nucleon
$ \Delta q^{i} = q^{i}_{+} -q^{i}_{-}$. Thus
in zero order of the perturbative QCD, i.e. in the limit of the free quark-parton
model the spin-dependent structure function $g_1$ can be expressed as
\cite{Jaffe:1987sx,Altarelli:1988nr}
\be\label{spinstructurefunction}
 g_1 = {1\over 2 } \sum_{i}  e^{2}_{i} [\Delta q^{i}(x,Q^2) + \Delta \bar{q}^{i}(x,Q^2) ].
\ee

 \begin{figure}
\includegraphics[width=9cm]{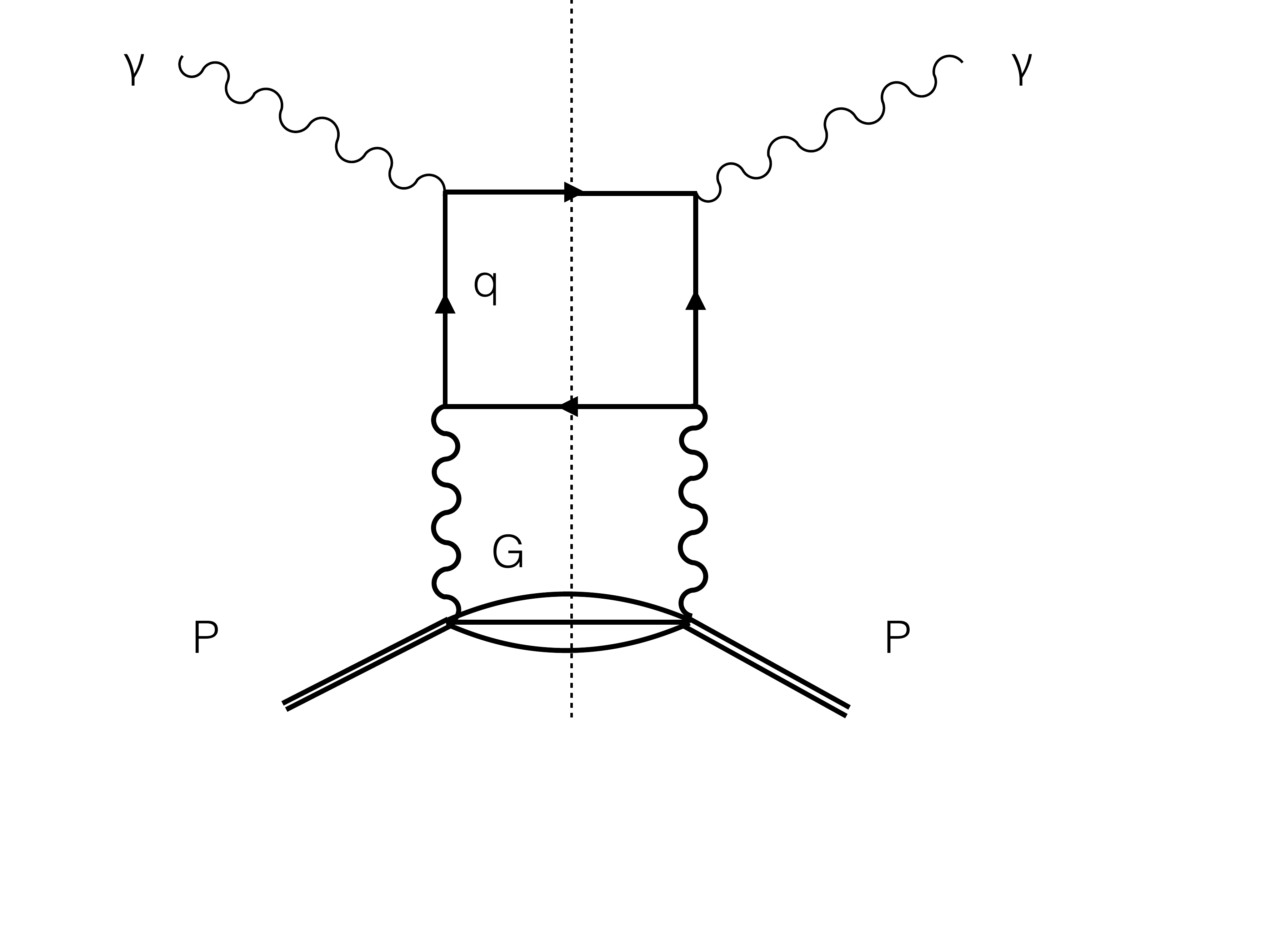}
 \centering
 \caption{The typical diagram contributing to the scattering of a photon on a proton gluon G  
 through the quark exchange.  }
 \label{fig2}
 \end{figure}

At large $Q^2$ the moments (\ref{structuremoments}) are given by the
OPE series with
the expansion coefficients related to nucleon matrix elements of the twist-2 operators
and Wilson coefficient functions in the form of QCD series in strong coupling constant
$\alpha = g^2/4\pi$.
The lowest ($n=1$) moments  of the spin-dependent nucleon structure functions
$g_{1}$  defined in (\ref{structuremoments}) can be represented in the following form
\cite{Kodaira:1978sh,Kodaira:1979ib,Kodaira:1979pa,Jaffe:1987sx,Altarelli:1988nr,
Close:1993mv,Larin:1997qq}:
\beqa\label{moments1}
\Gamma^{p}_{1} &=&  \int^{1}_{0} dx ~g^{p}_{1}(x,Q^2)= ~I_3 +I_8 +I_0\nn\\
\Gamma^{n}_{1}&=& \int^{1}_{0} dx ~g^{n}_{1}(x,Q^2)= - I_3 +I_8 +I_0,
\eeqa
where
\beqa\label{moments2}
I_3 &=&{1\over 12 } ~a_3~(1- {\alpha(Q^2) \over 2\pi}+...)\nn\\
I_8&=&{1 \over 36} ~a_8~(1- {\alpha(Q^2) \over 2\pi}+...)\nn\\
I_0&=&{1 \over 9} ~a_0~(1- {\alpha(Q^2) \over 2\pi}
{33-8 n_f \over 33- 2 n_f}+...).
\eeqa
The matrix elements of the non-singlet axial currents $J^k_{5\mu}$ and  the singlet axial current
$J_{5\mu}$ are  defined as follows\footnote{
The quark distributions are defined in terms of SU(3)-flavour
structure \cite{Aidala:2012mv,Anselmino:1994gn} :
$
 NS~ hypercharge~a_8 = \Delta u + \Delta d - 2 \Delta s,~
 NS ~ isovector~a_3 = \Delta u -  \Delta d,~
 Singlet~a_0 = \Delta u + \Delta d + \Delta s.
$
}:
\beqa\label{currents}
<p,s\vert \bar{\psi}  \gamma^{\mu}  \gamma_5  {\lambda^k \over 2} \psi  \vert p,s>&=&
M s^{\mu} ~  a_k,~~~~k=1,2,...,8\nn\\
<p,s\vert \bar{\psi}  \gamma^{\mu}  \gamma_5  \psi  \vert p,s> &=&
2 M s^{\mu}  ~ a_0.
\eeqa
The $\lambda^k$ are generators of the flavour group and $\psi =(u,d,s,...)$
is a vector  in flavour space.
The  $ a_k, a_0$ � are Lorentz invariant matrix elements and
reflect the unknown, nonperturbative aspect of the QCD dynamics.
The nucleon states $<p,s\vert$ are labeled by the momentum $p_{\mu}$ and the covariant spin vector
$s_{\mu}$ ($s \cdot p =0$, $s^2=-1$). There are perturbative corrections to the
coefficient functions which are included in expression (\ref{moments2}).

 \begin{figure}
\includegraphics[width=8cm]{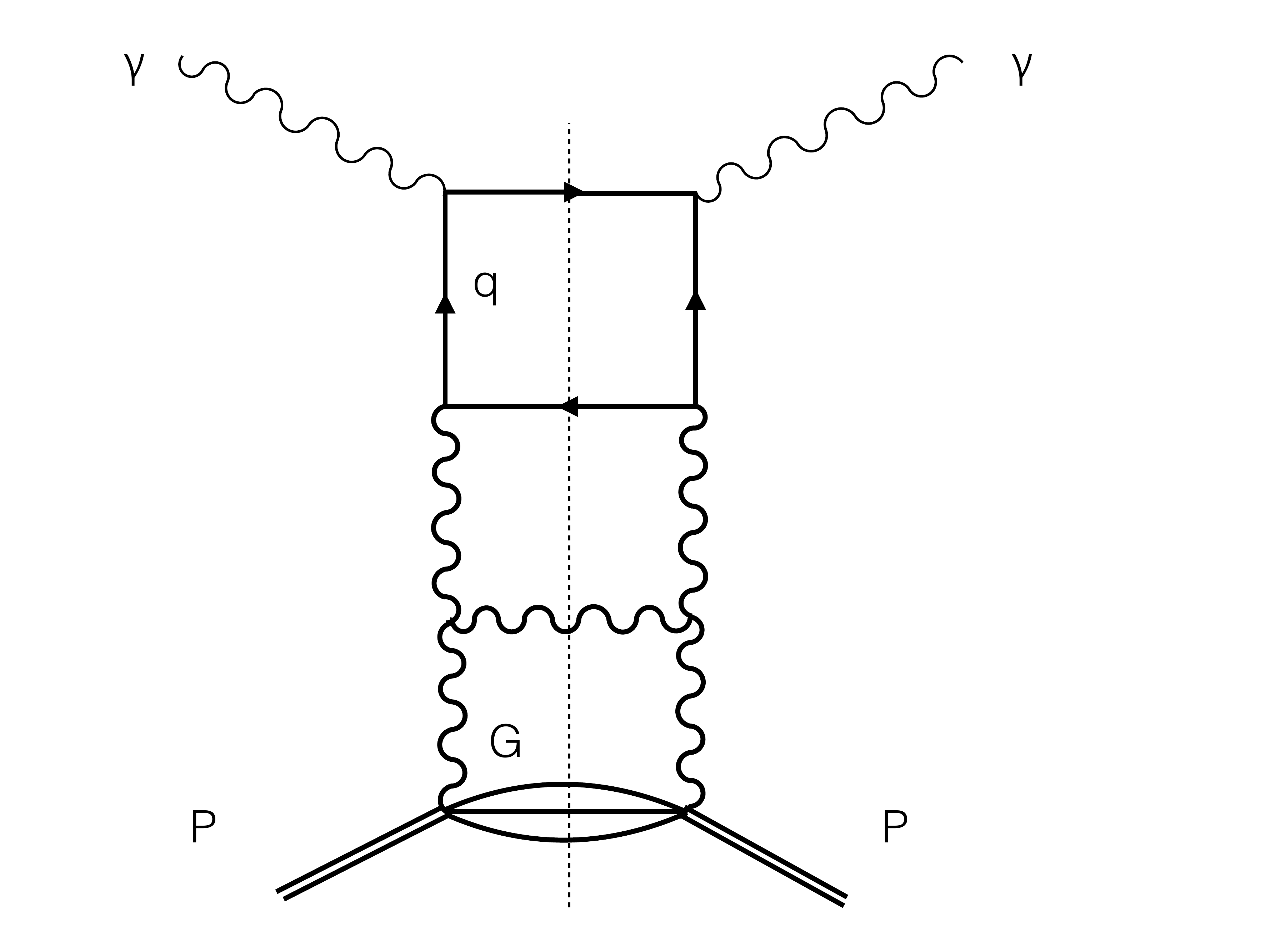}
 \centering
 \caption{Example of the diagram demonstrating the scattering of a
 photon on a proton gluon G in the  next to leading
 order through the quark and gluon exchange.  }
 \label{fig3}
 \end{figure}

To determine the flavour non-singlet components $a_3$ and $a_8$ on the right hand side of
the equations (\ref{moments1}), (\ref{moments2})  one should turn to the
beta-decay data \cite{Aidala:2012mv,Anselmino:1994gn,Close:1993mv,Close:1974ux}.
The values of  $a_3$ and $a_8$ extracted from hyperon
beta-decays under the assumption of SU(3) flavour symmetry
are\footnote{The invariant matrix elements $F$ and $D$ measured in beta-decay are:
$F=0.459 \pm 0.008$ and $D=0.798 \pm 0.008$.
\cite{Aidala:2012mv,Anselmino:1994gn}.~
}
\cite{Aidala:2012mv,Anselmino:1994gn,Close:1993mv,Close:1974ux,Bass:2009ed}:
\beqa\label{spinsumrule1}
a_8 &=& \Delta u + \Delta d - 2 \Delta s  = 3F-D~~~~~ ( 0.58 \pm 0.03)\nn\\
a_3&=&\Delta u - \Delta d = F+D~~~~~~~~~~~~~~~~ ( 1.270 \pm 0.003)
\eeqa
The experimental value of the quarks spin density $\Delta \Sigma =a_0$ can be obtained through the measurements of the first moment integrals of $g_1$ (\ref{moments1})
  $\Gamma^{p}_{1} = 0.135 \pm 0.011$, ~(\ref{moments2})
  $\Gamma^{n}_{1} = -0.028 \pm 0.006$,  of the strong coupling constant
 $\alpha_s =0.28 \pm 0.02$ all  at $<Q^2>=5GeV^2$   \cite{Close:1993mv,Alexakhin:2006oza}
and the  knowledge  of $a_3$ and $a_8$  in (\ref{spinsumrule1}), thus
\be\label{singaxialcurrent}
a_0= \Delta \Sigma =  \Delta u + \Delta d + \Delta s  = 0.33 \pm 0.03.
\ee
This value is unexpectedly small, compared with the naive
 expectation of $\Delta \Sigma_{SU(6)} \simeq 1$ and of the Ellis-Jaffe sum rule
$\Delta \Sigma_{EJ} \simeq 0.65$ \cite{Ellis:1973kp}.  By now it is experimentally well
established that indeed the matrix element $a_0$ of the flavour-singlet
axial-vector current is small and only of the order of $0.2-0.3$.
 The question raised was the following:
 why the quark spin content of the nucleon is so small?

 One of the possible answers considered in the literature consists in the realisation
that the singlet component $a_0$ receives an additional contribution from the gluon
polarisation $\triangle G$ \cite{Jaffe:1987sx,Efremov:1988zh,Altarelli:1988nr,Carlitz:1988ab,Lampe:1996mf}
\be
\Delta G = \int^{1}_{0} dx [G_+(x,Q^2) - G_-(x,Q^2)],
\ee
which is the amount of spin carried by polarised gluons
in the polarised  nucleon.

In general, the value of the matrix elements (\ref{currents}) depend on the renormalisation
scale $\mu^2$. Only if the operator is conserved, one can show that the
matrix elements are $\mu^2$ independent \cite{Gross:1974cs}. Because
the non-singlet currents $J^{3,8}_{5\mu}$ are exactly conserved in the massless limit,
the anomalous dimension of the non-singlet axial current vanishes at all orders. Thus
$ a_3$ and $a_8$ in (\ref{moments2}) are renormalisation group
invariant ($\mu^2$ independent), they have a real physics meaning and are
protected from substantial QCD radiative corrections \cite{Gross:1974cs}.
Perturbative corrections to the coefficient functions are presented  in (\ref{moments2})
\cite{Kodaira:1978sh,Kodaira:1979ib,Kodaira:1979pa,Jaffe:1987sx,Efremov:1988zh,Altarelli:1988nr,Carlitz:1988ab}.

The singlet axial current $J_{5\mu}= \bar{\psi}  \gamma_{\mu}  \gamma_5  \psi $
in (\ref{currents}) is not conserved \cite{Adler:1969gk,Bell:1969ts}:
\be\label{conservation}
\partial^{\mu} J_{5\mu} = n_f \ {\alpha \over 2\pi} \ tr G \wedge  G
= n_f \  {\alpha \over 2\pi} \ \partial^{\mu}   K_{\mu}.
\ee
Its expectation value $a_0(\mu^2)$ depends on the renormalisation  point $\mu^2$
and it is therefore not a physical quantity and  receives
anomalous contribution
\cite{Kodaira:1978sh,Kodaira:1979ib,Kodaira:1979pa,Jaffe:1987sx,Efremov:1988zh,Altarelli:1988nr,Carlitz:1988ab}.
Therefore the interpretation of the quark spin-density   $a_0 = \Delta \Sigma $  in
(\ref{moments2}), (\ref{currents})  and (\ref{singaxialcurrent})  as
the matrix element of the flavour singlet axial charge
was reconsidered \cite{Jaffe:1987sx,Efremov:1988zh,Altarelli:1988nr,Carlitz:1988ab}. The modified singlet axial  current
\be\label{anomaly}
 J_{5\mu} - n_f {\alpha  \over 2\pi} K_{\mu}
\ee
is conserved, as follows from (\ref{conservation}),
and its matrix element defines the invariant spin-density,
which  includes the polarisation of the gluons $\Delta G$ in the nucleon
\be
<p,s\vert J_{5\mu} - n_f {\alpha  \over 2\pi} K_{\mu}  \vert p,s> =
2 M s^{\mu}  ~ \hat{a}_0,
\ee
where now \cite{Altarelli:1988nr}
\be\label{gluonspinsscreening}
\hat{a}_0 ~ = ~\Delta \Sigma - n_f {\alpha  \over 2 \pi} \Delta G ~~ =~~ \Delta u + \Delta d + \Delta s
- n_f {\alpha  \over 2 \pi} \Delta G .
\ee

Thus the matrix elements of the flavor singlet axial vector current
are determined not solely by the first moment of the quark distributions,
but also by the first moment of a spin-dependent gluon distribution $\Delta G$.
This leads to possible explanations for the small value of  $\hat{a}_0$
extracted from polarised deep inelastic experiments (\ref{singaxialcurrent}) that
have been suggested in the literature \cite{Jaffe:1987sx,Altarelli:1988nr},
 which includes a
spin screening (\ref{gluonspinsscreening}) from positive
gluon polarisation $\Delta G$.
From the evolution equations it follows that the gluon polarisation
grow as
$$
\Delta G(Q^2) ~ ~\sim ~~  {const  \over   \alpha(Q^2) }
$$
at large $Q^2$. This corresponds to a scaling contribution $\alpha(Q^2) \Delta G(Q^2) = const$
to the first moment $I_{ep,en}$ (\ref{moments1})  of $g_1$
 (see formulas (\ref{scaling}) and (\ref{scaling1})).
 This suggestion sparked a   program
to measure $\Delta G $ which finally provides  the first evidence for nonzero
gluon polarisation in the proton \cite{Adolph:2012ca,Mallot:2006mc,Nocera:2012hx}.

 \begin{figure}
 \includegraphics[width=11cm]{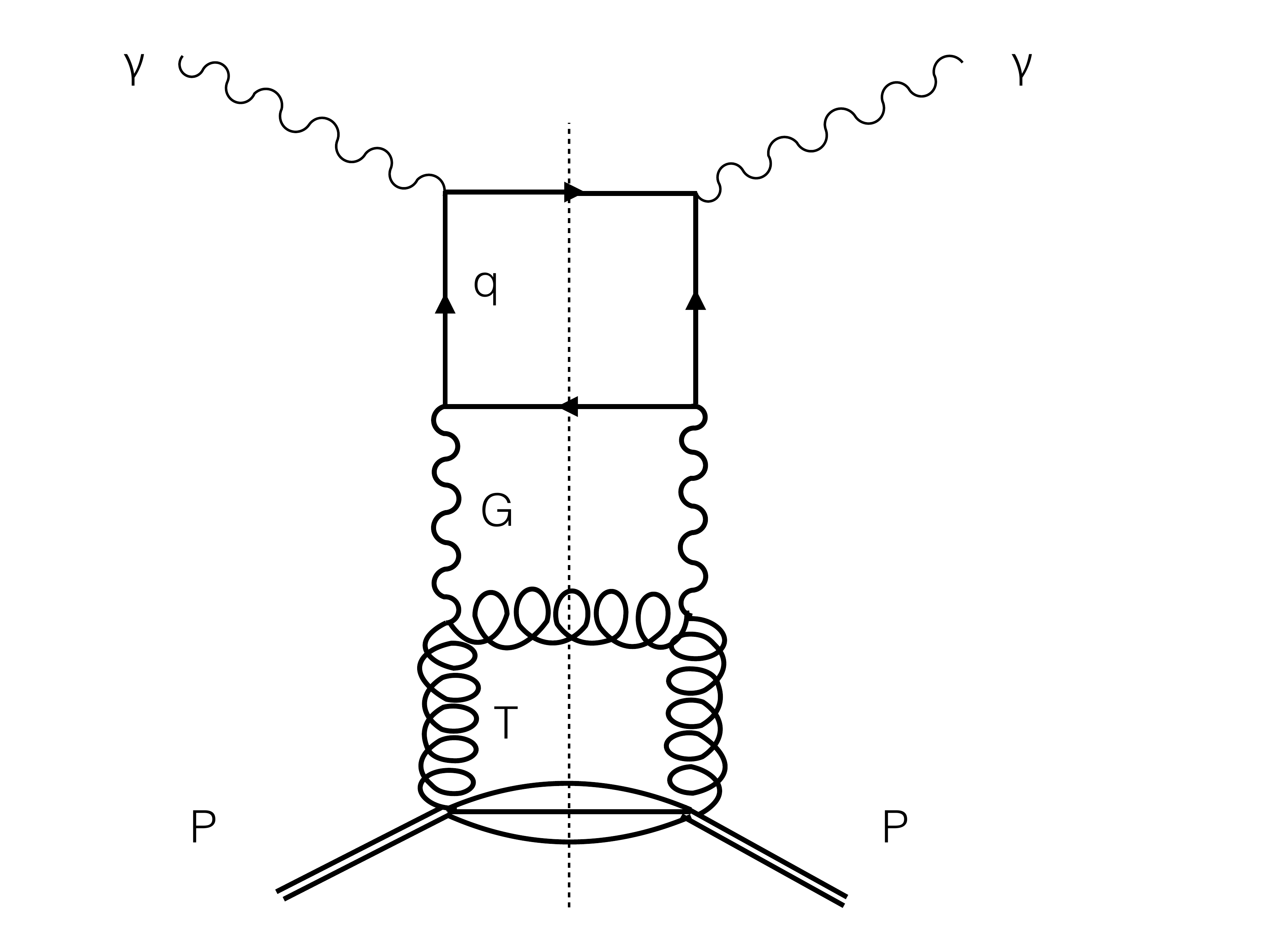}
 \centering
 \caption{Example of the diagram demonstrating the scattering of a
 photon on a proton tensorgluon T, which appears 
 in the next to leading  order through the quark and gluon exchange.}
 \label{fig4}
 \end{figure}

The polarised gluon distribution function $\Delta G $
was obtained from a QCD fit to the  data on
the spin-dependent structure function $g_1 $. The QCD fits
for $\Delta G $ were found in \cite{Adolph:2012ca,Mallot:2006mc,Nocera:2012hx}, one with
a positive and one with a negative first moment $\Delta G $, which describe the data equally well.
The absolute value is of the order of $\vert  \Delta G \vert \simeq 0.2-0.3$
for $Q^2 =3 GeV^2 $  and the uncertainty from the fit is of the
order of 0.1. {\it This amount of  gluon polarisation is not, by itself,
sufficient to resolve the problem of the small value of $\hat{a}_0$}. However that gluon
polarisation $\Delta G \approx 0.2 - 0.3$
would still make a significant contribution to the nucleon polarisation
 \cite{Adolph:2012ca,Mallot:2006mc,Nocera:2012hx,Ball:2013lla}.

Considering the influence of polarised tensorgluons density
$\Delta T_s$ to the spin of the nucleon  we found that $I_0 =  I^{'}_{0} +I^{''}_{0} $, the singlet
component of the first moments of $g_1$ in  (\ref{moments1})
 receives additional contribution $I^{''}_{0}$ from
 tensorgluons   (see Fig. \ref{fig4}):
\beqa\label{gluoncontribition}
 I^{'}_0&=&{1\over 9 }~(\Delta \Sigma - n_f {\alpha(Q^2) \over 2 \pi} \Delta G )~\left(1-{\alpha(Q^2)\over 2 \pi}
 {3 (12s^2-1)-8n_f\over 3 (12s^2-1)-2n_f} +...\right),
\eeqa
\beqa\label{tensorcontribition}
I^{''}_0&=&   n_f   {\alpha (Q^2)\over  2 \pi}  ~   \Delta T_s ~
{ \sum_{k=1}^{2s+1}{1\over k}\over 3 (12s^2-1)-2n_f} +...~.
\eeqa
In the absence of tensorgluons ( s=1) the $I^{'}_{0}$ in (\ref{gluoncontribition})
reduces to the standard QCD result (\ref{moments2}) and $I^{''}_{0}=0$ ($\Delta T =0$ ).
In the presence of the tensorgluons of spin s the result depends
on the spin factor $(12s^2-1)$,
which appears in the Callan-Simanzik beta function
coefficient \cite{Savvidy:2014hha,Savvidy:2013zwa}, as well
as on the contribution of the tensorgluon density $\Delta T_s$,
which scales similar to the quark density $\Delta \Sigma$ (see formulas
(\ref{tind}) and (\ref{sind})).
In general,  one should sum there over all "active" tensorgluon spins.
It is difficult to assist the effect of the tensorgluons  because it is not
known if $\Delta T_s$ is positive or negative. Note that if $\Delta T_s$ is negative,
i.e. if tensorgluons  give negative contribution to the proton helicity,
then the sign is correct for additional compensation of $\Delta \Sigma$ to occur.

In the next sections we shall derive evolution equations which include the
tensorgluon polarisation and their contribution to the singlet part of the
structure functions.

\section{\it Polarisation of Tensorgluons in Nucleon }

Our aim is to analyse a possible contribution of polarised tensorgluons $\Delta T_s$
into the spin of the nucleon if they are present in a nucleon. For that one should generalise the evolution equations for quark, gluons and tensorgluons derived in
\cite{Savvidy:2014hha,Savvidy:2013zwa} into the case of polarised tensorgluons.
The generalisation of the DGLAP evolution  equations which account for polarisation of
tensorgluons can be written in the following form:
\beqa\label{masterequation}
{dq^{i}_{+}(x,t)\over dt}&=&
{\alpha(t)\over 2\pi}\int_{x}^{1}{dy\over y}
\bigg[ q^{i}_{+}(y,t)P_{q_{+}q_{+}}({x\over y})+q_{-}^{i}(y,t)P_{q_{+}q_{-}}({x\over y}) + \nn\\
&&~~~~~~~~~~~~~~~~~~+
G_{+}(y,t)P_{q_{+}G_{+}}({x\over y})+G_{-}(y,t)P_{q_{+}G_{-}}({x\over y})
\bigg] \nn\\
{dq^{i}_{-}(x,t)\over dt}&=&
{\alpha(t)\over 2\pi}\int_{x}^{1}{dy\over y}
\bigg[ q^{i}_{+}(y,t)P_{q_{-}q_{+}}({x\over y})+q_{-}^{i}(y,t)P_{q_{-}q_{-}}({x\over y}) + \nn\\
&&~~~~~~~~~~~~~~~~~~+
G_{+}(y,t)P_{q_{-}G_{+}}({x\over y})+G_{-}(y,t)P_{q_{-}G_{-}}({x\over y})\bigg]\nn\\
{dG_{+}(x,t)\over dt}&=&
{\alpha(t)\over 2\pi}\int_{x}^{1}{dy\over y}
\bigg[ \sum_{i=1}^{2f}q^{i}_{+}(y,t)P_{G_{+}q_{+}}({x\over y})+\sum_{i=1}^{2f}q_{-}^{i}(y,t)P_{G_{+}q_{-}}({x\over y}) + \nn\\
&&~~~~~~~~~~~~~~~~~~+
G_{+}(y,t)P_{G_{+}G_{+}}({x\over y})+G_{-}(y,t)P_{G_{+}G_{-}}({x\over y})+ \nn\\
&&~~~~~~~~~~~~~~~~~~+
T_{+}(y,t)P_{G_{+}T_{+}}({x\over y})+T_{-}(y,t)P_{G_{+}T_{-}}({x\over y})
\bigg]\\
{dG_{-}(x,t)\over dt}&=&
{\alpha(t)\over 2\pi}\int_{x}^{1}{dy\over y}
\bigg[ \sum_{i=1}^{2f}q^{i}_{+}(y,t)P_{G_{-}q_{+}}({x\over y})+\sum_{i=1}^{2f}q_{-}^{i}(y,t)P_{G_{-}q_{-}}({x\over y}) + \nn\\
&&~~~~~~~~~~~~~~~~~~+
G_{+}(y,t)P_{G_{-}G_{+}}({x\over y})+G_{-}(y,t)P_{G_{-}G_{-}}({x\over y})+ \nn\\
&&~~~~~~~~~~~~~~~~~~+
T_{+}(y,t)P_{G_{-}T_{+}}({x\over y})+T_{-}(y,t)P_{G_{-}T_{-}}({x\over y})
\bigg] \nn\\
{dT_{+}(x,t)\over dt}&=&
{a(t)\over 2\pi}\int_{x}^{1}{dy\over y}
\bigg[G_{+}(y,t)P_{T_{+}G_{+}}({x\over y})+G_{-}(y,t)P_{T_{+}G_{-}}({x\over y})+ \nn\\
&&~~~~~~~~~~~~~~~~~~+
T_{+}(y,t)P_{T_{+}T_{+}}({x\over y})+T_{-}(y,t)P_{T_{+}T_{-}}({x\over y})
\bigg]\nn \\
{dT_{-}(x,t)\over dt}&=&
{\alpha(t)\over 2\pi}\int_{x}^{1}{dy\over y}
\bigg[G_{+}(y,t)P_{T_{-}G_{+}}({x\over y})+G_{-}(y,t)P_{T_{-}G_{-}}({x\over y})+ \nn\\
&&~~~~~~~~~~~~~~~~~~+
T_{+}(y,t)P_{T_{-}T_{+}}({x\over y})+T_{-}(y,t)P_{T_{-}T_{-}}({x\over y}),\nn
\eeqa
where $t=\ln(Q^2/Q^2_0)$.
This set of equations can be simplified by observing that parity conservation
in QCD implies the relations
\be
P_{B_{+}A_{\pm}}(z)=P_{B_{-}A_{\mp}}(z)
\ee
for any A and B and that the sums and the differences
\beqa\label{poladistributions}
 q^{i} = q^{i}_{+} + q^{i}_{-},~~~G = G_{+} + G_{-},~~~
 T = T_{+} + T_{-},\nn\\
\Delta q^{i} = q^{i}_{+} -q^{i}_{-},~~~\Delta G = G_{+} -G_{-},~~~
 \Delta T = T_{+} -T_{-}
\eeqa
evolve separately. Physically  the quantities  $\Delta q^{i},\Delta G ,
 \Delta T $ represent the differences
between the number densities of quarks, gluons
and tensorgluons with spin parallel to the nucleon
spin  and those with spin anti-parallel.
It is also convenient to define the sum and the difference
of splitting functions:
\be
P_{BA}(z)=P_{B_{+}A_{+}}(z)+P_{B_{-}A_{+}}(z), ~~~~~
\Delta P_{BA}(z)=P_{B_{+}A_{+}}(z)-P_{B_{-}A_{+}}(z).
\ee
We derive then from (\ref{masterequation}) the master equation for differences:
\beqa\label{nextmaster}
&{d \over dt}\Delta q^{i}(x,t)=
{\alpha(t)\over 2\pi} \int_{x}^{1} {dy\over y}
\bigg[ \Delta q^{i}(y,t) \Delta P_{q q}({x\over y})
 +
\Delta G(y,t) \Delta P_{q G}({x\over y})
\bigg] \nn \\
 \nn\\
&   {d\over dt} \Delta G(x,t)=
{\alpha(t)\over 2\pi}\int_{x}^{1}{dy\over y}
\bigg[ \sum_{i=1}^{2n_f} \Delta q^{i}(y,t) \Delta P_{G q}({x\over y})
 +
 \Delta G(y,t) \Delta P_{G G}({x\over y})   +
 \Delta T(y,t) \Delta P_{G T}({x\over y})
\bigg] \nn\\
&{d\over dt} \Delta T(x,t)=
{\alpha(t)\over 2\pi}\int_{x}^{1}{dy\over y}
\bigg[ \Delta G(y,t) \Delta P_{T G}({x\over y})  +
 \Delta T(y,t) \Delta P_{T T}({x\over y})
\bigg].
\eeqa
In the  case of non-singlet (NS) quark densities the above equation
reduces to the one of the standard QCD:
\beqa
{d \over dt}\Delta q^{NS}(x,t)=
{\alpha(t)\over 2\pi} \int_{x}^{1} {dy\over y}
  \Delta q^{NS}(y,t) \Delta P_{q q}({x\over y}),
\eeqa
where $q^{NS}(y,t)= q(x,t)- \bar{q}(x,t)$.
The splitting function
$P_{q_- q_+}=0$ and therefore $\Delta P_{qq}= P_{q_+ q_+}=P_{qq}$. It follows then that
the moments of $\Delta q^{NS}$  evolve in $Q^2$ with the same logarithmic
exponent as $q^{NS}$ and therefore as the non-singlet parts of $F_1$ and $g_1$.

The integral
\be
\Delta q^{i} = \int^{1}_{0} dx     [\Delta q^{i}(x,t) + \Delta \bar{q}^{i}(x,t) ]
\ee
represents the fraction of the proton's spin carried by the quarks and
the anti-quarks  of the flavour $q^{i}$. The summation over all quark flavour
contributions gives the total
 {\it fraction of the nucleon spin carried by light quark and antiquarks}
(below the contribution of the heavy quarks has been neglected):
\be
\Delta \Sigma  =  \int^{1}_{0} dx \sum_i[\Delta q^{i}(x,t) + \Delta \bar{q}^{i}(x,t) ]=
 \Delta u + \Delta d +\Delta s .
\ee

As we already reviewed in the introduction, the conservation of the SU(3) singlet axial current,
valid for massless quarks, is broken in QCD by the Adler-Bell-Jackiw anomaly
\cite{Adler:1969gk,Bell:1969ts}.
While the anomalous
dimension of the non-singlet axial current vanishes at all orders
\cite{Kodaira:1978sh,Kodaira:1979ib},
the singlet axial current anomalous dimension is different from zero.
Due to this non-conservation
in the singlet sector there is  a gluon contribution
at all values of $Q^2$ in the singlet part of the integral of $g_1$.
This gluon term $\Delta G$ in (\ref{gluonspinsscreening})
\be
\hat{a}_0= \Delta \Sigma - n_f {\alpha  \over 2 \pi} \Delta G
\ee
can in principle be large and we are interested to
consider a possible contribution of the tensorgluons polarisation as well.
With this aim we shall introduce in addition to the gluon polarisation the  corresponding
tensorgluon helicity content in the polarised nucleon:
\be
\Delta G = \int^{1}_{0} dx [G_+(x,Q^2) - G_-(x,Q^2)], ~~~~
\Delta T = \int^{1}_{0} dx [T_+(x,Q^2) - T_-(x,Q^2)],
\ee
with $G_{\pm}, T_{\pm}$ being the gluon and tensorgluons densities with helicities
$\pm 1$ and $\pm s$ respectively in a nucleon with helicity $+{1\over 2}$.
$\Delta G$ represents the amount of helicity carried by the gluons in the nucleon and
$\Delta T_s$, the amount carried by the tensorgluons.

The nucleon's spin is accounted by the quarks'  spin as well as  by gluon and tensorgluon
spins $\Delta G$,  $\Delta T_s$ and the total orbital angular
momentum $L_z$.
Because the spin of the nucleon is $1/2$ one should have
\be
{1\over 2}  \Delta \Sigma + \Delta G + \sum_{s} s ~\Delta T_s  + L_z  ={1\over 2}.
\ee
In this equation the helicity of the gluon is equal to one and of tensorgluon
is $s=2,3,..$, and
the summation is over all "active" tensor bosons in the nucleon.
The various components of the nucleon spin in the above
equation can be measured experimentally.  The gluon sector $\Delta G$ can be measured in deep inelastic scattering experiments in which gluon-gluon
or gluon-quark scattering dominates. One of the
most interesting suggestions to come out of the spin crisis
is that $\Delta G$ may contribute significantly to the nucleon. Our
aim is to consider a possible contribution of the tensorgluon spin $s \triangle T_s$.

\section{\it Evolution Equation in Leading and Next to Leading Order}

At leading
order the QCD evolution of $\Delta \Sigma$, $\Delta G$ and $\Delta T$ can
written from (\ref{nextmaster}) as

\beqa\label{evolution3}
{d \over d t} \left(\begin{array}{c}
\Delta \Sigma \\
\Delta G \\
\Delta T
\end{array} \right)= {\alpha(t)\over 2\pi}
\left(\begin{array}{ccc}
 \Delta \gamma_{qq}^{(1)}& 2 n_f \Delta \gamma_{qG}^ {(1)}&0\\
 \Delta \gamma_{Gq}^ {(1)}& \Delta \gamma_{GG}^ {(1)}& \Delta \gamma_{GT}^ {(1)}\\
 0&\Delta \gamma_{TG}^ {(1)}&\Delta \gamma_{TT}^ {(1)}
\end{array} \right)
\left(\begin{array}{c}
\Delta \Sigma \\
\Delta G \\
\Delta T
\end{array} \right),
\eeqa
where \cite{Savvidy:2013zwa} (see Appendix A for the values of the corresponding anomalous dimensions)
\beqa\label{anomamartix1}
 \Delta \gamma_1= \left(\begin{array}{ccc}
0& 0&0\\
 {3\over 2} C_2(R)&  C_2(G) {\sum_s(12s^2-1)\over 6} -{2\over 3}  n_f T(R) &
 C_2(G) \sum^{2s+1}_{k= 1}
{1\over  k } \\
 0&0&0
\end{array} \right),
\eeqa
and  $C_2(G)= N, C_2(R)={N^2-1  \over  2 N},  T(R) = {1  \over  2}$ for the SU(N) groups.
Let us represent the $\triangle \gamma_1$ as
\beqa
\Delta \gamma_1&=&\left(
\begin{array}{ccc}
0 & 0 & 0\\
C_{1} & b & C_{3}\\
0 & 0 & 0 \end {array}\right)\nn,
\eeqa
where \cite{Savvidy:2013zwa}
\be\label{betafunction}
C_{1}={3\over 2}C_{2}(R),~~
b= C_2(G) {\sum_s(12s^2-1)\over 6} -{2\over 3}  n_f T(R),~~
 C_{3}=C_2(G) \sum^{2s+1}_{k= 1}{1\over  k }
\ee
and
\be\label{couplingevol}
{d \alpha(t) \over d t}   = - {b \over 2 \pi}   \alpha^2(t).
\ee
The eigenvalues  of the matrix $\Delta \gamma_1$ are
$
 \lambda_1 =b,~\lambda_{2,3}=0
$
and the corresponding eigenvectors
\beqa
S=\left(
\begin{array}{ccc}
-{C_{3}\over C_{1}} & -{b\over C_{1}} & 0\\
0 & 1 & 1\\
1 & 0 & 0 \end {array}\right),~~~~~~
S^{-1} = \left(
\begin{array}{ccc}
0 &0 & 1\\
-{C_{1}\over b} & 0 & -{C_{3}\over b}\\
{C_{1}\over b} & 1 & {C_{3}\over b} \end {array}\right)\nn
\eeqa
define the transformation
$
 \Delta \gamma_1 = S\Lambda S^{-1},
$
where $\Lambda = diag(0,0,b)$, which
allows  to represent the evolution equation (\ref{evolution3}) in the following form:
\beqa
{d\over dt}\left(
\begin{array}{c}
{\Delta \Sigma}\\
{\Delta G}\\
{\Delta T}\end {array}\right)
={\alpha(t) \over 2\pi}S\Lambda S^{-1}
\left(
\begin{array}{c}
{\Delta \Sigma}\\
{\Delta G}\\
{\Delta T}\end {array}\right).
\eeqa
Multiplying this equation from the left by $S^{-1}$
we shall get
\beqa
{d\over dt}\bigg[
S^{-1}\left(
\begin{array}{c}
{\Delta \Sigma}\\
{\Delta G}\\
{\Delta T}\end {array}\right)\bigg]=
{\alpha(t)\over 2\pi}\Lambda \bigg[ S^{-1}
\left(
\begin{array}{c}
{\Delta \Sigma}\\
{\Delta G}\\
{\Delta T}\end {array}\right)\bigg]\nn,
\eeqa
and it is useful to introduce the new coordinates
\beqa\label{evol}
\left(
\begin{array}{c}
{y_{1}(t)}\\
{y_{2}(t)}\\
{y_{3}(t)}\end {array}\right)=S^{-1}\left(
\begin{array}{c}
{\Delta \Sigma}\\
{\Delta G}\\
{\Delta T}\end {array}\right)=
\left(
\begin{array}{c}
{\Delta T}\\
{-{1\over b}(C_{1} \Sigma+C_{3}\Delta T)}\\
{{C_{1}\over b}  \Sigma+\Delta G+{C_{3}\over b}\Delta T}
\end {array}\right).
\eeqa
The evolution equation takes a diagonal form:
\beqa
{d\over dt}
\left(
\begin{array}{c}
{y_{1}(t)}\\
{y_{2}(t)}\\
{y_{3}(t)}\end {array}\right)=
{\alpha(t)\over 2\pi}
\left(
\begin{array}{ccc}
0 & 0 & 0\\
0 & 0 & 0\\
0 & 0 & b \end {array}\right)
\left(
\begin{array}{c}
{y_{1}(t)}\\
{y_{2}(t)}\\
{y_{3}(t)}\end {array}\right).
\eeqa
The first equation gives
$
{dy_{1}(t)\over dt}={d\Delta T(t)\over dt}=0,
$
and we conclude that the helicity density of tensorgluons
does  not evolve  with $Q^2$:
\be\label{tind}
\Delta T(t)=\Delta T  .
\ee
From this and from the second equation
${dy_{2}(t)/ dt}=0$ it follows that the quark helicity distribution
also does  not evolve  with $Q^2$
($
{d \Delta \Sigma(t) / dt}=0
$):
\be\label{sind}
\Delta  \Sigma(t)= \Delta \Sigma .
\ee
From the last equation
\beqa
{dy_{3}(t)\over dt}&=&{\alpha(t) \over 2\pi}~ b ~y_{3}(t)
\eeqa
it follows then that
\beqa
{d ~ \alpha(t)  ~y_{3}(t)\over dt}&=& {d  \alpha(t)  \over dt} ~y_{3}(t) +
\alpha(t) {d   y_{3}(t)\over dt} = \nn\\
&=& - {b \over 2 \pi}   \alpha^2(t) ~y_{3}(t) +
\alpha(t) {\alpha(t)\over 2\pi}~ b ~y_{3}(t)=0,
\eeqa
where we use (\ref{couplingevol}). Thus
\be
\alpha(t) ~ y_{3}(t) = Const
\ee
and therefore from the last equation in (\ref{evol})
\be\label{spin}
{\alpha(t) \over 2 \pi } \Delta G(t) =Const-  {\alpha(t) \over 2 \pi   b}
[ C_{3} \Delta T  +  C_{1} \Delta  \Sigma]  \rightarrow Const.
\ee
as $t \rightarrow \infty$ and $\alpha(t)  \rightarrow 0$. Here we use the fact that
$\Delta \Sigma$ and $\Delta T$  are t-independent (\ref{tind}), (\ref{sind}).
In the absence of tensorgluons the equation reduces to the Altarelli-Ross equation:
\be\label{scaling}
{\alpha(t) \over 2 \pi } \Delta G(t) =Const -  {\alpha(t) \over 2 \pi   b}
  C_{1}  \Delta \Sigma \rightarrow Const.
\ee
Therefore it is natural to introduce the quantity \cite{Altarelli:1988nr}
\be\label{scaling1}
\Delta\Gamma = {\alpha(t) \over 2 \pi } \Delta G(t)
\ee
and consider its behaviour in the next to leading order.
One should notice that the coefficient in front of the tensorgluon density $C_3$
in (\ref{spin}) grows with spin as
$
  \ln s
$,
therefore its contribution  can still be numerically large for higher spins.
{\it The main conclusion which follows from the above consideration is that the densities
$\Delta \Sigma$ and $\Delta T$ are $Q^2$ independent and that the gluon
density grows with $Q^2$ as $\Delta G \approx 1/\alpha(Q^2)$.}

Expanding the splitting functions and anomalous dimensions into  perturbative series
\beqa
\Delta P_{ij}= \Delta P^{(1)}_{ij}+{a_{s}(t)\over 2\pi} \Delta P^{(2)}_{ij}+\Big({a_{s}(t)\over 2\pi}\Big)^2 \Delta P^{(3)}_{ij}+\dots
\eeqa
one can get from (\ref{nextmaster}):
\beqa\label{nnnmaster}
{d\over dt}\left(\begin{array}{c}
\Delta \Sigma(t)\\
\Delta\Gamma(t)\\
\Delta T(t)
\end{array} \right)=
\Big({a_{s}(t)\over 2\pi}\Big)^2
\left(\begin{array}{ccc}
\Delta \gamma_{qq}^{(2)} & 2n_f  \Delta \gamma_{qG}^{(3)}&0\\
\Delta \gamma_{Gq}^{(1)}&\Delta \gamma_{GG}^{(2)}&\Delta \gamma_{GT}^{(1)}\\
0&\Delta \gamma_{TG}^{(3)}&\Delta \gamma_{TT}^{(2)}\
\end{array} \right)
\left(\begin{array}{c}
\Delta \Sigma(t)\\
\Delta\Gamma(t)\\
\Delta T(t)
\end{array} \right).
\eeqa
Here we have  taken into account that there are no direct interactions
between tensorgluons and quarks
$
\gamma_{Tq}^{(2)}=\gamma_{qT}^{(2)}=0
$
\cite{Savvidy:2014hha,Savvidy:2013zwa}.
The $Q^2$ evolution of the singlet component of the first moment  (\ref{moments1})
is given by the short-distance operator product expansion
\cite{Kodaira:1978sh,Kodaira:1979ib,Kodaira:1979pa,Altarelli:1988nr}:
\beqa\label{singlet}
I_0&=&{1\over 9}c(a_{s}(t))E(a_{s}, a_{s}(t)) \CO (a_{s}),
\eeqa
where we take into account the presence of the tensorgluons:
\beqa\label{vectors}
c(a_{s}(t))&=&\Big(1+c_{\Sigma}a_{s}(t), ~~c_{\Gamma}+c_{\Gamma}^{'}a_{s}(t),
~~c_{T}+c_{T}^{'}a_{s}(t)\Big),~~~
\CO(a_{s})=\left(\begin{array}{c}
\Delta \Sigma(t)\\
\Delta\Gamma(t)\\
\Delta T(t)
\end{array} \right)
\eeqa
and
\beqa\label{matrixofanomalies}
E(a_{s}, a_{s}(t))&=&1+{a_{s}-a_{s}(t)\over 4 \pi b}~ \Delta \gamma~,
\eeqa
$\gamma$ is the matrix of anomalous dimensions in equation (\ref{nnnmaster}).
Substituting the matrix of anomalous dimensions  (\ref{matrixofanomalies}) ,
(\ref{nnnmaster}) and the vectors (\ref{vectors})
into the (\ref{singlet}) we shall get
\beqa\label{singlet1}
I_0
&=&{1\over 9}\bigg\{
\bigg[
1+{a_{s}-a_{s}(t)\over 4 \pi b}\Big(\Delta\gamma_{qq}^{(2)}
+c_{\Gamma} \Delta \gamma_{Gq}^{(1)}
-bc_{\Sigma}\Big)+c_{\Sigma}a_{s}\bigg]  \Delta \Sigma+\nn \\
&&~~+
\bigg[
1+{a_{s}-a_{s}(t)\over 4 \pi b}\Big( \Delta \gamma_{GG}^{(2)}
+{1\over c_{\Gamma}} \Delta \gamma_{qG}^{(3)}+
{c_{T}\over c_{\Gamma}} \Delta \gamma_{TG}^{(3)}-b{c_{\Gamma}^{'}\over c_{\Gamma}}\Big)+{c_{\Gamma}^{'}\over c_{\Gamma}}a_{s}\bigg] c_{\Gamma}
\Delta\Gamma+\nn \\
&&~~~+
\bigg[
c_{T} + {a_{s}-a_{s}(t)\over 4 \pi b}   \Big(  c_{T} \Delta \gamma_{TT}^{(2)}+
 c_{\Gamma} \Delta \gamma_{GT}^{(1)}-
b c_{T}^{'} \Big)
+ c_{T}^{'}  a_{s}\bigg]    \Delta T\bigg\}.
\eeqa
One can see the appearance of the tensorgluon contribution $\Delta T$ into the first  moment
of the singlet polarised structure function.  In this expression
the anomalous dimensions associated with the quark and gluon interactions
were computed in \cite{Kodaira:1978sh,Kodaira:1979ib,Kodaira:1979pa,Altarelli:1988nr}. These values are
$
\Delta \gamma_{qq}^{(2)}=0
$
and
\beqa
c_{\Gamma}&=&-n_f,~~
c_{\Sigma}=-{3\over 4\pi}C_{2}(R),~~
c'_{\Gamma}=c_{\Gamma}c_{\Sigma}\nn \\
\Delta \gamma_{Gq}^{(1)}&=& {3\over 2}C_{2}(R),~~
\Delta \gamma_{qG}^{(3)}=c_{\Gamma}^{2} \Delta \gamma_{Gq}^{(1)},~~
\Delta \gamma_{GG}^{(2)}=0.
\eeqa
Because there are no  interactions between photon  and tensorgluons
through the quark exchanges,  the coefficients $c_{T}$ and $c'_{T}$  vanish.
Hence, keeping terms proportional to $\alpha(t)$ and using the  above values for the anomalous dimensions,
the first moment of the singlet part of the proton structure function (\ref{singlet1})
takes the following form:
\beqa
 I^{'}_0 &=&{1\over 9}
\bigg[1-{a_{s}(t)\over 4 \pi b}(c_{\Gamma} \Delta \gamma_{Gq}^{(1)}
-bc_{\Sigma})\bigg]
(\Delta \Sigma - n_f \Delta\Gamma)\nn\\
 I^{''}_0&=&-
{1\over 9} \bigg[{a_{s}(t)\over 4 \pi b}c_{\Gamma} \Delta \gamma_{GT}^{(1)}\bigg]~\Delta T  ,\nn
\eeqa
where $I_0 = I^{'}_0+I^{''}_0$.
The coefficient of the beta function $b$ of the running coupling constant is given in (\ref{betafunction}).
Given that $C_{2}(G)=3$, $C_{2}(R)={4\over 3}$ and $T(R)={1\over 2}$
we finally get:
\beqa
 I^{'}_0&=&{1\over 9 }~(\Delta \Sigma - n_f {\alpha_s(t) \over 2 \pi} \Delta G ) ~\left(1-{a_{s}(t)\over 2 \pi}
 {3\sum(12s^2-1)-8n_f\over 3\sum(12s^2-1)-2n_f} +...\right)
\eeqa
\beqa
I^{''}_0&=&    n_f \sum_s          {a_{s}(t)\over  2 \pi} ~ \Delta T_s ~
{ \sum_{k=1}^{2s+1}
{1\over k}\over 3\sum(12s^2-1)-2n_f}  +...
\eeqa
Comparing this result with the standard QCD case one can see that the assumed presence
of the tensorgluons inside the proton modifies the singlet part of the polarised proton
structure function in two ways.  On the one hand, there is
a contribution $3\sum(12s^2-1)$ in the numerator and the denominator
to the radiative correction $I^{'}_0$  which effectively lowers its value.
On the other hand, there is an additional contribution $I^{''}_0$  of the
tensorgluons  $\Delta T_s$. The overall sign of this contribution depends on the values of the spin $s$ of the tensorgluon  and of its density $\Delta T_s$. Note that if $\Delta T_s$ is negative,
i.e. if tensorgluons  give negative contribution to the proton helicity,
then the sign is correct for the additional compensation of $\Delta \Sigma$ to occur.

\section{\it Acknowledgements}
On of us (G.S.) would like to thank  Guido Altarelli and Yury Dokshitzer for discussions 
and Pavel Galumian for pointing out to him the nucleon spin issue.
This work was supported in part by the
General Secretariat for Research and Technology of Greece and
from the European Regional Development Fund MIS-448332-ORASY (NSRF 2007-13 ACTION, KRIPIS).

\section{\it Appendix A}
Since $P_{B_{+}A_{\pm}}(z)=P_{B_{-}A_{\mp}}(z)$, we only need to evaluate
$P_{B_{+}A_{+}}(z)$, $P_{B_{-}A_{+}}(z)$. Then
\[P_{BA}(z)=P_{B_{+}A_{+}}(z)+P_{B_{-}A_{+}}(z), ~~~~~
\Delta P_{BA}(z)=P_{B_{+}A_{+}}(z)-P_{B_{-}A_{+}}(z).\]
Hence for the $\triangle P_{qq}$  \cite{Altarelli:1977zs}:
\beqa
P_{q_{-}q_{+}}(z)&=&0
\nn \\
\triangle P_{qq}= P_{q_{+}q_{+}}(z)= P_{qq}= &=&C_{2}(R)\bigg[ {1+z^{2}\over (1-z)_{+}}
    +{3 \over 2}\delta(z-1)\bigg],
\eeqa
for the $\triangle P_{Gq}(z)$
\beqa
P_{G_{+}q_{+}}(z)=C_{2}(R){1\over z},~~& &~~
P_{G_{-}q_{+}}(z)=C_{2}(R){(1-z)^{2}\over z}\nn\\
P_{Gq}(z)=P_{G_{+}q_{+}}(z)+P_{G_{-}q_{+}}(z)&=&
C_{2}(R) {1+(1-z)^{2}\over z} \nn\\
\triangle P_{Gq}(z)= P_{G_{+}q_{+}}(z)-P_{G_{-}q_{+}}(z)&=&
C_{2}(R) {1-(1-z)^{2}\over z},
\eeqa
for the $\triangle P_{qG}(z)$
\beqa
P_{q_{+}G_{+}}(z)={z^2\over 2},~~& &~~
P_{q_{-}G_{+}}(z)={(z-1)^2\over 2}\nn \\
P_{qG}(z)=P_{q_{+}G_{+}}(z)+P_{q_{-}G_{+}}(z)&=&{  z^2+(z-1)^2\over 2} \nn\\
\triangle P_{qG}(z)=P_{q_{+}G_{+}}(z)-P_{q_{-}G_{+}}(z)&=&{z^2-(z-1)^2 \over 2}
\eeqa
as well as for the $\triangle P_{GG}(z)$
\beqa
P_{G_{+}G_{+}}(z)&=&C_{2}(G)\bigg[
(1+z^{4})\bigg({1\over z}+{1\over (1-z)_{+}}\bigg)+\bigg({\sum_s(12s^2-1)\over 6}
-{2\over 3}{T(R)\over C_{2}(G)}\bigg)
\delta(z-1)\bigg]\nn \\
P_{G_{-}G_{+}}(z)&=&C_{2}(G){(1-z)^{3}\over z}\nn \\
\triangle P_{G G }(z)&=&C_{2}(G)\bigg[
(1+z^{4})\bigg({1\over z}+{1\over (1-z)_{+}}\bigg) -{(1-z)^{3}\over z}
+\bigg({\sum_s(12s^2-1)\over 6}
-{2\over 3}{T(R)\over C_{2}(G)}\bigg)
\delta(z-1)\bigg]\nn.
\eeqa
The behaviour at small $x \rightarrow 0$ of the unpolarised and polarised splitting functions 
is different  
\beqa\label{smalx}
\lim_{x \rightarrow 0}P_{GG}(z)=C_{2}(G) 
 \bigg( {1\over z}+{1\over  z }\bigg) ,~~~~~~
\lim_{x \rightarrow 0}\triangle P_{G G }(z)=C_{2}(G) 
 \bigg({1\over z}- {1\over  z }\bigg)  =0.
\eeqa
The polarised splitting function $\triangle P_{G G }(z)$ has moderate  behaviour at small x's.
The polarised splitting functions which are including tensorgluons were found
in \cite{Savvidy:2014hha,Savvidy:2013zwa}. For the non-diagonal functions they are:
\beqa
P_{T_{+}G_{+}}(z)&=&C_{2}(G)   {z^{2s+1}\over  (1-z)^{2s-1}}  \nn \\
P_{T_{-}G_{+}}(z)&=&C_{2}(G) {(1-z)^{2s+1}\over z^{2s-1}}  \nn \\
\triangle P_{TG}(z)&=&
C_{2}(G)\bigg[
{z^{2s+1}\over  (1-z)^{2s-1}}
-  {(1-z)^{2s+1}\over z^{2s-1}}   \bigg],
\eeqa
and at small x both functions are singular
\beqa
\lim_{x \rightarrow 0} P_{TG}(z)=C_{2}(G)   {1\over   z^{2s-1} },~~~~~~~~~
\lim_{x \rightarrow 0} \triangle P_{TG}(z)= - C_{2}(G) 
{1 \over z^{2s-1}} .
\eeqa
Comparing this behaviour with the gluon splitting functions (\ref{smalx}) one can 
see that here the polarised splitting function has enhanced singular behaviour in 
the small x region. Finally for the $P_{TG}$ function we have
\beqa
P_{G_{+}T_{+}}(z)&=&C_{2}(G){1\over z(1-z)^{2s-1}} \nn\\
P_{G_{-}T_{+}}(z)&=&C_{2}(G){ (1-z)^{2s+1}\over z } \nn \\
\triangle P_{GT}(z) &=&
C_{2}(G)\bigg[{1\over z(1-z)^{2s-1}}-
{ (1-z)^{2s+1}\over z }\bigg].
\eeqa
In the small x region we have behaviour which is similar to gluon  (\ref{smalx})
\beqa
\lim_{x \rightarrow 0} P_{GT}(z)=C_{2}(G)   {2\over   z },~~~~~~~~~
\lim_{x \rightarrow 0} \triangle P_{GT}(z)= 0.
\eeqa
The tensor-tensor splitting function is helicity conserving:
\beqa
P_{T_{-}T_{+}}(z)&=&0 \\
\triangle P_{TT}(z)=P_{T_{+}T_{+}}(z)&=&C_{2}(G)\bigg[
{z^{2s+1}\over  (1-z)_{+}} +{1\over (1-z) z^{2s-1}_{+}}+
\sum_{j=1}^{2s+1}{1\over j}\delta(z-1)\bigg] \nn
\eeqa
and is singular in the small x region both functions are singular
\beqa
\lim_{x \rightarrow 0} P_{TT}(z)=C_{2}(G)   {1\over   z^{2s-1} },~~~~~~~~~
\lim_{x \rightarrow 0} \triangle P_{TT}(z)= C_{2}(G) 
{1 \over z^{2s-1}} .
\eeqa

The matrix of anomalous dimensions is \cite{Altarelli:1977zs}:
\beqa\label{anomamartix}
 \int^{1}_{0} dz z^{n-1}\left(\begin{array}{ccc}
  \triangle P_{qq}(z)&2 n_f \triangle P_{qG}(z)&0\\
  \triangle P_{Gq}(z)& \triangle P_{GG}(z)&P_{GT}(z)\\
 0&\triangle P_{TG}(z)& \triangle P_{TT}(z)
\end{array} \right)=
 \left(\begin{array}{ccc}
 \triangle \gamma_{qq}(n)& 2 n_f \triangle \gamma_{qG}(n)&0\\
 \triangle \gamma_{Gq}(n)& \triangle \gamma_{GG}(n)& \triangle \gamma_{GT}(n)\\
 0&\triangle \gamma_{TG}(n)&\triangle \gamma_{TT}(n) \nn
\end{array} \right),
\eeqa
where
\beqa\label{anomamartixqqG}
 \int^{1}_{0} dz 
  \triangle P_{qq}(z)&=&0,~~~~~~~~~n=1,\nn\\
\int^{1}_{0} dz 
  \triangle P_{qG}(z)&=& 0,~~~~~~~~ \nn\\
  \int^{1}_{0} dz 
  \triangle P_{Gq}(z)&=&  {3\over 2}C_2(R),~~~~~~~~   \nn\\
\int^{1}_{0} dz 
  \triangle P_{GG}(z)&=&  {\sum_s(12s^2-1)\over 6} C_2(G)
 - {2 \over 3}T(R),~~~~~~~~ \nn
\eeqa
and for the tensorgluons \cite{Savvidy:2014hha,Savvidy:2013zwa}
\beqa\label{anomamartixGT}
\int_{0}^{1} dz   \triangle P_{GT}(z) &=&  C_2(G) \sum^{2s+1}_{k= 1}
{1\over  k },~~~~ n= 1, \\
\int_{0}^{1} dz  \triangle P_{TG}(z) &=&  0,\nn\\
\int_{0}^{1} dz  \triangle P_{TT}(z) &=& 0.\nn
\eeqa

\vfill
\end{document}